\documentclass[aps,tightenlines]{revtex4}%

\usepackage{amsfonts}
\usepackage{amsmath}
\usepackage{amssymb}
\usepackage{graphicx}

\begin{document}

\title{Multipartite entanglement of three trapped ions in a cavity and W-
State generation}
\author{S. Shelly Sharma}
\email{shelly@uel.br}
\affiliation{Depto. de F\'{\i}sica, Universidade Estadual de Londrina, Londrina
86051-990, PR Brazil }
\author{Eduardo de Almeida}
\affiliation{Depto. de F\'{\i}sica, Universidade Estadual de Londrina, Londrina
86051-990, PR Brazil }
\author{Naresh K. Sharma}
\affiliation{Depto. de Matematica, Universidade Estadual de Londrina, Londrina 86051-990,
PR Brazil }

\begin{abstract}
A scheme to generate three qubit maximally entangled W-states, using three
trapped ions interacting with red sideband tuned single mode field of a high
finesse cavity, is proposed. For the cavity field initially prepared in a
number state, the probability of generating three ion W-state is calculated.
By using the ion-cavity coupling strengths achieved in experimental
realizations, the interaction time needed for W-state generation is found to
be of the order of 10 $\mu $ sec. It is found that for a fixed number of
photons in the cavity the nature of entanglement of ionic internal states
can be manipulated by appropriate choice of initial state phonon number.

The ionic qubits in W-like state are found to be entangled to cavity
photons. Analytical expressions for global negativity and partial $K-$way
negativities ($K=2$ to $4$) are obtained to study the evolution of
entanglement distribution as a function of interaction parameter. Reversible
entanglement exchange between different entanglement modes is observed. For
specific values of interaction parameter, the three ions and photon-phonon
system are found to have four partite entanglement, generated by $2-$way and 
$3-$way correlations.
\end{abstract}

\maketitle

\section{Introduction}

Interest in manipulation of multipartite quantum states is motivated by the
possible use in quantum communication \cite{eker91}, quantum dense coding 
\cite{benn92} and quantum teleportation \cite{benn93}. Advances in
experimental techniques of trapping, cooling, and manipulation of internal
states of ions through interaction with external lasers \cite%
{monr95,wine98,roos99} have made the generation of maximally entangled
states \cite{haff05} a reality. While single atoms and ions, with long-lived
internal states, are suitable for storing quantum information, the photonic
qubits serve as fast and reliable carriers to transport quantum information
over long distances. Cirac et al. \cite{cira97} considered the possibility
of using photonic channels to interconnect high finesse cavities with single
trapped ions inside. It was shown in ref. \cite{buze97}, that the composite
quantum system with a single trapped two-level ion interacting with a
quantized light field in single-mode cavity evolves into maximally entangled
three-particle Greenberger-Horne-Zeilinger state. A different scheme to
generate three qubit maximally entangled GHZ state, using a trapped ion
interacting with a resonant external laser and sideband tuned single mode of
a cavity field has been proposed in ref. \cite{shar03} and the decoherence
process of this three qubit system examined \cite{shar06}. Experimentally
successful efforts in this direction \cite{mund02,mund03,esch03}, have
demonstrated coherent coupling of electronic and motional states of a single
trapped Ca$^{+}$ ion to single field mode of a high finesse cavity. In the
experiment of ref. \cite{mund02}, a Ca$^{+}$ ion in a RF-Paul trap was
placed with high precision at an arbitrary position in the standing wave
field of a near-confocal resonator for several hours of interaction time.
Since the size of the cavity is large compared to the wavelength of standing
wave, a large number of nodes and antinodes of the standing wave are formed
inside the cavity. The electromagnetically induced transparency (EIT) method
used in the work of Morigi et. al. \cite{mori00} \ and Roos et al. \cite%
{roos00}, could make possible the initialization of motional states of ion
strings in a cavity. These experiments open the possibility that more than
one trapped ions may be cooled down to their lowest vibrational states and
interact with the quantized cavity field. A string of ions in the same trap
with the center of mass in a definite motional mode may, likewise, be
prepared inside a cavity. In this article, we propose a scheme in which
three two-level trapped ions are cooled inside a high finesse cavity to
generate multipartite entangled states through coherent coupling of their
electronic and vibrational states to the standing wave cavity field. For the
cavity field initially prepared in a number state, we obtain conditions on
the interaction parameters to generate states in which three ions in W like
state are entangled to cavity photons. Tessier et al. \cite{tess03} have
investigated the multipartite entanglement of two two-level atoms, generated
by resonant coupling to a single mode of the electromagnetic field. The
resulting physical system corresponds to the two-atom Tavis-Cummings model 
\cite{tavi68}. Our system is similar to that of ref. \cite{tess03} in that
the entanglement between the ions arises not through direct interaction
between the ions, but through their coupling to the phonon-photon system. We
have not considered the effects of spontaneous decay, decoherence of
motional states and cavity losses in this article. The object of studying a
closed system, without decoherence effects, is to have a clear picture of
dynamics of quantum correlations. The entanglement of mixed multipartite
states resulting due to decoherence is not so well understood. The memory
qubits (ionic internal states) are known to have a coherence time of the
order of seconds and even minutes (magnetically insensitive transitions).
The success of experiments with a single ion indicates, that the motional
decoherence caused by trap fluctuations or environment noise is negligible
on the time-scale of state generation and photon qubit coherence of the
order of milisecs. The decoherence caused by the implementation of the
scheme itself needs to be studied theoretically for the three ion case
though such studies for single ion case have been done in refs. \cite%
{fidi02,shar06}.

Multipartite entanglement is a useful resource for implementation of quantum
information processing protocols \cite{bouw00}. To retrieve information
about the entanglement of a group of $K$ $(2\leq K\leq 4)$ subsystems of the
composite system, we use the global negativity and the partial $K-$way
negativities. {For analyzing the entanglement dynamics, we have considered
the vibrational} {phonons and cavity photons to constitute a single quantum
system. }Negativity \cite{zycz98,vida00,eise01}, based on Peres Horodecki 
\cite{pere96,horo96} criterion, has been shown to be an entanglement
monotone \cite{vida02}. The $K-$way negativity refers to the negativity of a
partial transpose constructed by imposing specific constraints during
transposition. The coherences of a multipartite composite system having N
subsystems can be quantified by $K-$way negativities \cite{shar06b} ($2\leq
K\leq N)$. Partial $K-$way negativity is the contribution of a specific $K-$%
way partial transpose to global negativity. For canonical states, the
partial $K-$way negativities measure the genuine $K-$partite entanglement of
the system \cite{shar07,shar08}. Genuine $K-$partite entanglement refers to $%
K-$partite entanglement due to correlations simlar to those present in a
GHZ-like state of $K$ subsystems. Loss of a single subsystem destroys this
type of entanglement completely, leaving no residual entanglement. We have
obtained analytical expressions for partial $K-$way negativities with
respect to photon-phonon state and internal states of ions. The advantage of
using partial $K-$way negativities lies in the fact that the entanglement
between parts of a composite system is obtained from the full state
operator, without state reduction. With the cavity prepared initially in a
Fock state, the entanglement of ionic internal states depends strongly on
the choice of phonon number at $t=0$. Extra control on qubit state
manipulation, gained by coupling the vibrational modes to the cavity field 
\cite{mund03,esch03}, is an advantage for successful implementation of
quantum gates.

The vector space and system Hamiltonian are discussed in section II.
Analytical expressions for the state of the system at current time, with the
three ions prepared in their respective (i) ground states, and (ii) excited
states, are presented in section III. The role of initial state center of
mass quanta in state manipulation and the time evolution of W-state
generation probabilities is also discussed in section III. In section IV,
the entanglement dynamics of ions, phonons and photons is investigated. The
Global and partial $K-$way negativities, used to study the entanglement
distribution in the composite system, are defined and calculated
analytically as well as numerically for special initial state preparations
of the system. We discuss, briefly, the information gained from the dynamics
of global and partial $K-$way negativities in section V, followed by the
conclusions in section VI.

\section{The Model}

Consider three two level trapped cold ions vibrating with trap frequency $%
\nu $ inside a high finesse cavity. The frequency of standing wave cavity
field is $\omega _{c}$ and ions are well separated from each other so that
no dipole interaction takes place between the ions. The free Hamiltonian of
the system composed of internal states of ions, the vibrational state of
center of mass and the state of standing wave cavity field is given by

\begin{equation}
\hat{H}_{0}=\hbar \nu \left( \hat{a}^{\dagger }\hat{a}+\frac{1}{2}\right)
+\hbar \omega _{c}\hat{b}^{\dagger }\hat{b}+\frac{\hbar \omega _{0}}{2}%
\sum_{j=1}^{3}\hat{\sigma}_{z}^{(j)},  \label{1}
\end{equation}%
where $\hat{a}^{\dagger }(\hat{a})$ and $\hat{b}^{\dagger }(\hat{b})$ are
the creation(annihilation) operators for vibrational phonons and cavity
field photons, respectively. Eigen states of Pauli operator $\hat{\sigma}%
_{z}^{(j)}$ model the internal states of the $j^{th}$ two-level ion ($%
j=1,2,3 $) with energy splitting $\hbar \omega _{0}$. We define the total
spin operators as $\hat{\sigma}_{k}=\sum_{j=1}^{3}\hat{\sigma}_{k}^{(j)}$,
where $k=(z,+,-)$ and use the eigenvectors of $\widehat{\sigma }^{2}$ and $%
\hat{\sigma}_{z}$ to represent the three ion internal states. The ionic \
internal states \ in product basis are labelled as $\left\vert \sigma
_{z}^{(1)}\sigma _{z}^{(2)}\sigma _{z}^{(3)}\right\rangle $ ($\sigma
_{z}^{(j)}=-1$ or $+1$))$.$ The coupled basis vectors are $|\sigma ,\sigma
_{z}\rangle $, where the label $\sigma $($=2s$) refers to the eigenvalue of $%
\sigma ^{2}$ given by $\sigma (\sigma +2)$. The computational basis states,
on the other hand, read as $\left\vert i_{1},i_{2},i_{3}\right\rangle ,$
where $i_{j}=0$ for atom in ground state and $i_{j}=1$ for atom in the
excited state.

The interaction of cold ions, located at the node of the standing wave, with
the quantized cavity field is given by

\begin{equation}
\hat{H}_{I}=\hbar g(\hat{\sigma}_{+}+\hat{\sigma}_{-})(\hat{b}^{\dagger }+%
\hat{b})\sin [\eta (\hat{a}^{\dagger }+\hat{a})],  \label{2}
\end{equation}%
$g$ being the ion-cavity coupling constant and $\eta $ the Lamb-Dicke
parameter. The interaction picture Hamiltonian obtained by applying the
unitary transformation $\hat{U_{I}}=\exp \left( -i\hat{H}_{0}t/\hbar \right) 
$ is a complex looking operator \cite{shar03b}. However, with the cavity
coupled to red sideband of ionic vibrational motion ($\omega _{0}-\omega
_{c}=\nu $ ), the relevant part of $\hat{H}_{I}$ in the rotating-wave
approximation and Lamb-Dicke limit ($\eta \ll 1)$ reduces to

\begin{equation}
\hat{H}_{II}=\hbar g\eta \lbrack \hat{\sigma}_{+}\hat{b}\hat{a}+\hat{\sigma}%
_{-}\hat{b}^{\dagger }\hat{a}^{\dagger }].  \label{3}
\end{equation}%
The possible values of $\sigma $ for three ions are $3,1,1$, there being two
distinct internal configurations that allow $\sigma =1$. The matrix $T$ that
transforms from the computational basis to the coupled basis is given by%
\begin{equation}
T=\left( 
\begin{array}{cccccccc}
1 & 0 & 0 & 0 & 0 & 0 & 0 & 0 \\ 
0 & \frac{1}{\sqrt{3}} & \frac{1}{\sqrt{3}} & 0 & \frac{1}{\sqrt{3}} & 0 & 0
& 0 \\ 
0 & 0 & 0 & \frac{1}{\sqrt{3}} & 0 & \frac{1}{\sqrt{3}} & \frac{1}{\sqrt{3}}
& 0 \\ 
0 & 0 & 0 & 0 & 0 & 0 & 0 & 1 \\ 
0 & \frac{1}{\sqrt{6}} & \frac{1}{\sqrt{6}} & 0 & -\sqrt{\frac{2}{3}} & 0 & 0
& 0 \\ 
0 & 0 & 0 & \sqrt{\frac{2}{3}} & 0 & \frac{-1}{\sqrt{6}} & \frac{-1}{\sqrt{6}%
} & 0 \\ 
0 & \frac{1}{\sqrt{2}} & \frac{-1}{\sqrt{2}} & 0 & 0 & 0 & 0 & 0 \\ 
0 & 0 & 0 & 0 & 0 & \frac{1}{\sqrt{2}} & \frac{-1}{\sqrt{2}} & 0%
\end{array}%
\right) .  \label{4}
\end{equation}%
The computational basis vectors are taken in the order {$|000\rangle $, $%
|100\rangle $, $|010\rangle $, $|110\rangle $,$|001\rangle $, $|101\rangle $%
, $|011\rangle $, and $|111\rangle $}, while the ordering of coupled basis
vectors is {$|3,-3\rangle $, $|3,-1\rangle $, $|3,1\rangle $, $|3,3\rangle $%
, $|1,-1\rangle _{1}$, $|1,1\rangle _{1}$, $|1,-1\rangle _{2}$, $|1,1\rangle
_{2}$}. The subscript in $|1,\pm 1\rangle _{1,2}$ distinguishes the states
with same value of $\sigma $ but different internal configurations.

The product state of the composite system looks like {$|\sigma ,\sigma
_{z}\rangle \otimes $}$\left\vert m,n\right\rangle $, where $m$ is the
number of center of mass vibrational quanta and $n$ is the number of cavity
field photons. The advantage of working in a coupled basis stems from the
fact that the Hamiltonian of Eq. (\ref{3}) does not connect states with
different values of $\sigma $. If the initial state is an eigen state of $%
\hat{\sigma}^{2}$ with eigenvalue $\sigma _{I}(\sigma _{I}+2)$, the system
evolves into a linear combination of states of the type $|\sigma _{I},\sigma
_{z},m,n\rangle $ with $-\sigma _{I}\leq \sigma _{z}\leq \sigma _{I}$.

{For }$\sigma _{I}${$=3$ the Hamiltonian (Eq. (\ref{3})) in the basis\ $%
|3,3\rangle \otimes |m-2,n-2\rangle $, $|3,1\rangle \otimes |m-1,n-1\rangle $%
, $|3,-1\rangle \otimes |m,n\rangle $ , and $|3,-3\rangle \otimes
|m+1,n+1\rangle $ is written as }

\begin{equation}
H_{_{II}}({\sigma =3})=\left( 
\begin{array}{cccc}
0 & \sqrt{2}A_{mn} & 0 & 0 \\ 
\sqrt{2}A_{mn} & 0 & \sqrt{2}B_{mn} & 0 \\ 
0 & \sqrt{2}B_{mn} & 0 & \sqrt{2}C_{mn} \\ 
0 & 0 & \sqrt{2}C_{mn} & 0%
\end{array}%
\right) ,
\end{equation}%
where 
\begin{equation}
A_{mn}=\hbar g\eta \sqrt{\frac{3}{2}(m-1)(n-1)},B_{mn}=\hbar g\eta \sqrt{2mn}%
,
\end{equation}%
and 
\begin{equation}
C_{mn}=\hbar g\eta \sqrt{\frac{3}{2}(m+1)(n+1)}.
\end{equation}%
Defining 
\begin{equation*}
\mu =(A_{mn}^{2}+B_{mn}^{2}+C_{mn}^{2}),\text{ and \ }\beta =\sqrt{\mu
^{2}-4A_{mn}^{2}C_{mn}^{2}},
\end{equation*}%
the eigenvalues of $H$, in units of $\hbar g\eta $ , are 
\begin{eqnarray}
E_{1} &=&-\sqrt{\mu -\beta },\qquad E_{2}=\sqrt{\mu -\beta },  \notag \\
E_{3} &=&-\sqrt{\mu +\beta },\qquad E_{4}=\sqrt{\mu +\beta }.
\end{eqnarray}%
The unitary matrix $\hat{U}$ that diagonalizes $H_{_{II}}({\sigma =3})$, $%
\left( H_{_{II}}({\sigma =3})|\phi _{i}\rangle =E_{i}|\phi _{i}\rangle ,\
(i=1,4)\right) $ is given by 
\begin{equation}
\hat{U}({\sigma =3})=\left( 
\begin{array}{cccc}
\sqrt{\frac{\beta +\mu _{2}}{4\beta }} & -\sqrt{\frac{\beta -\mu _{1}}{%
4\beta }} & -\sqrt{\frac{\beta -\mu _{2}}{4\beta }} & \sqrt{\frac{\beta +\mu
_{1}}{4\beta }} \\ 
-\sqrt{\frac{\beta +\mu _{2}}{4\beta }} & -\sqrt{\frac{\beta -\mu _{1}}{%
4\beta }} & \sqrt{\frac{\beta -\mu _{2}}{4\beta }} & \sqrt{\frac{\beta +\mu
_{1}}{4\beta }} \\ 
-\sqrt{\frac{\beta -\mu _{2}}{4\beta }} & \sqrt{\frac{\beta +\mu _{1}}{%
4\beta }} & -\sqrt{\frac{\beta +\mu _{2}}{4\beta }} & \sqrt{\frac{\beta -\mu
_{1}}{4\beta }} \\ 
\sqrt{\frac{\beta -\mu _{2}}{4\beta }} & \sqrt{\frac{\beta +\mu _{1}}{4\beta 
}} & \sqrt{\frac{\beta +\mu _{2}}{4\beta }} & \sqrt{\frac{\beta -\mu _{1}}{%
4\beta }}%
\end{array}%
\right) ,
\end{equation}%
where $\mu _{1}=\mu -2A_{mn}^{2}$ and $\mu _{2}=\mu -2C_{mn}^{2}.$

Starting from a given initial state, $\left\vert \sigma _{z}^{(1)}\sigma
_{z}^{(2)}\sigma _{z}^{(3)}\right\rangle \otimes \left\vert m,n\right\rangle 
$, of the system, the time evolution in interaction picture is obtained by
applying the transformation $\exp \left( -i\hat{H}_{II}t/{\hbar }\right) $.
An internal state of ions may be rewritten in terms of the coupled basis
states as

\begin{equation}
\left\vert \sigma _{z}^{(1)},\sigma _{z}^{(2)},\sigma
_{z}^{(3)}\right\rangle _{j}=\sum_{i}T_{ji}^{\dagger }|\sigma ,\sigma
_{z}\rangle _{i}.  \label{10}
\end{equation}%
The state $\left\vert \sigma ,\sigma _{z}\right\rangle \otimes $ $\left\vert
m,n\right\rangle _{i}$ of the composite system with $m$ vibrational quanta
and $n$ photons may, in turn, be expanded in terms of eigen states of $%
H_{II}(\sigma )$ as

\begin{equation}
\left\vert \sigma ,\sigma _{z}\right\rangle \otimes \left\vert
m,n\right\rangle _{i}=\sum_{k}U_{ik}^{\dagger }(\sigma )|\phi _{k}\rangle .
\label{11}
\end{equation}%
We recall that the matrix $T$ operates on the internal states of the
two-level atom, while $\hat{U}$ operates in the composite Hilbert space
formed by internal states, cavity field states and the state of vibrational
motion. The state of the system at instant $t$ is given by

\begin{equation}
\Psi (t)=\sum_{i}\sum_{k}T_{ji}^{\dagger }U_{ik}^{\dagger }(\sigma )\exp
\left( {-i}\lambda _{k}g\eta t\right) |\phi _{k}\rangle .  \label{12}
\end{equation}%
To go back to the computational basis we use the inverse transformation. The
state obtained is in the interaction picture and we go to Schr\"{o}dinger
picture by the transformation $\Psi _{S}(t)=e^{-\frac{i\hat{H_{0}}t}{\hbar }%
}\Psi (t)$. As the state of the composite system evolves, the internal
states of ions, the cavity field state and the state of vibrational motion
become entangled.

\section{The composite system state}

\subsection{Initial state, $\Psi _{m+1,n+1}(0)=\left\vert 000\right\rangle
\otimes |m+1,n+1\rangle $}

Consider the three ions in their ground states with the center of mass
motion cooled down to a state with $m+1$ vibrational quanta. Photon field
set to a fixed detuning is injected in to the cavity such that the cavity
state is an $n+1$ photon Fock state at $t=0$. The analytical expression for
the state of the system at current time $t$ is found to be%
\begin{eqnarray}
\Psi _{m+1,n+1}(t) &=&a_{0}(t)|000\rangle \otimes |m+1,n+1\rangle
+a_{1}(t)|W_{1}\rangle \otimes |m,n\rangle   \notag \\
&&+a_{2}(t))|W_{2}\rangle \otimes |m-1,n-1\rangle +a_{3}(t))|111\rangle
\otimes |m-2,n-2\rangle   \label{13}
\end{eqnarray}%
where $\left\vert W_{1}\right\rangle =\frac{|100\rangle +|001\rangle
+|010\rangle }{\sqrt{3}}$ and $\left\vert W_{2}\right\rangle =\frac{%
|110\rangle +101\rangle +|011\rangle }{\sqrt{3}}$ are $W$ states with one
and two excited ions, respectively. The coefficients $a_{i}(t),i=0$ to $3$,%
\begin{equation}
a_{0}(t)=\frac{e^{-i\omega _{1}t}}{2\beta }\left( (\beta -\mu _{2})\cos {%
\left( \sqrt{(\mu +\beta )}g\eta t\right) }+(\beta +\mu _{2})\cos {\left( 
\sqrt{(\mu -\beta )}g\eta t\right) }\right) {,}  \label{14}
\end{equation}%
\begin{eqnarray}
a_{1}(t) &=&-\frac{ie^{-i\omega _{1}t}}{2\beta }\left( \sqrt{(\beta -\mu
_{2})(\beta +\mu _{1})}\sin {\left( \sqrt{(\mu +\beta )}g\eta t\right) }%
\right.   \notag \\
&&\left. +\sqrt{(\beta +\mu _{2})(\beta -\mu _{1})}\sin {\left( \sqrt{(\mu
-\beta )}g\eta t\right) }\right) ,  \label{15}
\end{eqnarray}%
\ \ 
\begin{equation}
a_{2}(t)=\frac{e^{-i\omega _{1}t}\sqrt{\beta ^{2}-\mu _{2}^{2}}}{2\beta }%
\left( \cos {\left( \sqrt{(\mu +\beta )}g\eta t\right) }-\cos {\left( \sqrt{%
(\mu -\beta )}g\eta t\right) }\right) ,  \label{16}
\end{equation}%
and%
\begin{eqnarray}
a_{3}(t) &=&-\frac{ie^{-i\omega _{1}t}}{2\beta }\left( {\sqrt{(\beta -\mu
_{2})(\beta -\mu _{1})}\sin {\left( \sqrt{(\mu +\beta )}g\eta t\right) }}%
\right.   \notag \\
&&\left. {-\sqrt{(\beta +\mu _{2})(\beta +\mu _{1})}\sin {\left( \sqrt{(\mu
-\beta )}g\eta t\right) }}\right) ,  \label{17}
\end{eqnarray}%
satisfy the normalization condition $\sum_{i=0}^{3}\left\vert
a_{i}(t)\right\vert ^{2}=1$. The frequency $\omega _{1}=\nu (m+\frac{3}{2}%
)+\omega _{c}(n+1)-\frac{3\omega _{0}}{2},$ refers to zero point energy of
the initial state. The probability amplitude $a_{2}(t)$ is zero whenever $%
\cos {\left( \sqrt{(\mu +\beta )}g\eta t\right) =}\cos {\left( \sqrt{(\mu
-\beta )}g\eta t\right) .}$ For values of $t$ such that $\cos {\left( \sqrt{%
(\mu +\beta )}g\eta t\right) =}\cos {\left( \sqrt{(\mu -\beta )}g\eta
t\right) =\pm 1}$, the composite system state is a separable state. When the
condition $\sin {\left( \sqrt{(\mu +\beta )}g\eta t\right) =}\sin {\left( 
\sqrt{(\mu -\beta )}g\eta t\right) =\pm 1}$ is satisfied, the composite
system is found to be in the state \ 
\begin{equation}
\Psi _{m+1,n+1}^{W_{1}}(t)=a_{1}(t)|W_{1}\rangle \otimes |m,n\rangle
+a_{3}(t))|111\rangle \otimes |m-2,n-2\rangle ,  \label{psiw1}
\end{equation}%
where the probability amplitudes 
\begin{equation}
a_{1}(t)=-\frac{ie^{-i\omega _{1}t}}{2\beta }\left( \sqrt{(\beta -\mu
_{2})(\beta +\mu _{1})}+\sqrt{(\beta +\mu _{2})(\beta -\mu _{1})}\right) ,
\label{a1}
\end{equation}%
\begin{equation}
a_{3}(t)=-\frac{ie^{-i\omega _{1}t}}{2\beta }\left( {\sqrt{(\beta -\mu
_{2})(\beta -\mu _{1})}-\sqrt{(\beta +\mu _{2})(\beta +\mu _{1})}}\right) ,
\label{a3}
\end{equation}%
depend strongly on the initial state photon and phonon number. The three
ions in $W_{1}$- like state are found to be entangled to photon-phonon
state, constituting a state having four-partite entanglement.

The probabilities $P_{i}(\tau )$ of finding $\ i$ number of ions ($i=0$ to $%
3 $) in excited state are plotted as a function of variable $\tau $ in
figure (\ref{fig1}), for the choice $m=n=2$. The value of parameter $\tau
=g\eta t$ is determined by the cavity ion coupling strength, Lamb Dicke
parameter and the interaction time. We notice that for $\tau \approx 3\pi /4$
the system is found to be in a separable state and for $\tau \approx 3\pi /8$
in state $\Psi _{3,3}^{W_{1}}$, with $P_{1}(\tau )\approx 0.75$, and $%
P_{3}(\tau )\approx 0.25$.

\subsection{Initial state, $\Phi _{m-2,n-2}(0)=)|111\rangle \otimes
|m-2,n-2\rangle $}

When all three ions are in their excited states at $t=0,$ the center of mass
prepared in a state with $m-2$ vibrational quanta, and the cavity in $n-2$
photon Fock state, the state of the composite quantum system at current time
is found to be%
\begin{eqnarray}
\Phi _{m-2,n-2}(t) &=&a_{3}(t))|000\rangle \otimes |m+1,n+1\rangle
+a_{2}(t))|W_{1}\rangle \otimes |m,n\rangle  \label{23} \\
&&+a_{1}(t))|W_{2}\rangle \otimes |m-1,n-1\rangle +a_{0}(t))|111\rangle
\otimes |m-2,n-2\rangle .  \notag
\end{eqnarray}%
The maximum number of coupled basis states populated by the ions as the
interaction time increases is four, independent of the initial state photon
or phonon number. Recalling that for interaction time such that $\sin {%
\left( \sqrt{(\mu +\beta )}g\eta t\right) =}\sin {\left( \sqrt{(\mu -\beta )}%
g\eta t\right) =\pm 1}$, the probability amplitude $a_{0}(t)=a_{2}(t)=0,$ we
find the three ions in $W_{2}$-like state entangled to photon-phonon state
such that%
\begin{equation}
\Phi _{m-2,n-2}^{W_{2}}(t)=a_{3}(t))|000\rangle \otimes |m+1,n+1\rangle
+a_{1}(t))|W_{2}\rangle \otimes |m-1,n-1\rangle ,  \label{psiw2}
\end{equation}%
with probability amplitudes $a_{1}(t)$ and $a_{3}(t)$ given by Eqs(\ref{a1})
and (\ref{a3}) respectively. Figure (\ref{fig2}) displays the probabilities $%
P_{i}(\tau )$ ($i=0$ to $3)$ as a function of variable $\tau $ for the
choice $m=2,n=2$ at $t=0$. The initial state preparation in this case
involves cooling the center of mass motion to zero phonon mode with ions in
ground state and the cavity in vacuum state. The internal states of three
ions can now be prepared in the excited state through controlled resonant
interaction with external lasers.\ The ion's state is well preserved due to
the long lifetime ($\sim $1 s). We observe that the composite system
periodically returns to initial separable state with a period of $t\approx 
\frac{3\pi }{4g\eta }$. The black arrows point out the $P_{2}(\tau )$
corresponding to state $\Phi _{00}^{W_{2}}(t).$

\subsection{Quantum state control and the number of vibrational quanta}

The maximum number of coupled basis states populated by the ions can be
controlled by the number of initial state photon and phonon number. With the
center of mass prepared initially in one phonon mode ($m+1=1$), and the
cavity having one or more photons ($n+1\geq 1$), the composite system is
found to be in the state 
\begin{align}
\Psi _{1,n+1}(t)& =\cos {\left( \sqrt{3(n+1)}g\eta t\right) }e^{-i\omega
_{1}t})|000\rangle \otimes |1,n+1\rangle   \notag \\
& -i\sin {\left( \sqrt{3(n+1)}g\eta t\right) }e^{-i\omega
_{1}t})|W_{1}\rangle \otimes |0,n\rangle ,  \label{18}
\end{align}%
whereas for ionic center of mass in zero phonon mode, the state of the
system remains unchanged. The minimum interaction time needed to get the
three ion $W_{1}$ state generation probability peak is ${t}_{\min }$ ${=}%
\frac{\pi }{2g\eta \sqrt{3(n+1)}}${. Deterministic coupling of ion's
quantized vibration in the trap }to the cavity mode, has been demonstrated
by Mundt et al. \cite{mund03}. {For cavity ion coupling strength }$g=8.95$
MHz, Lamb Dicke parameter value $\eta =0.01$ and {cavity prepared in single
photon state at }$t=0$, we obtain $t_{\min }=\frac{\pi }{2g\eta \sqrt{3}}%
=10.133\mu $ sec. This is the simplest setting for\ generating three ion $W$
state with a single ion in excited state. The three ion $|W_{1}\rangle $
state generation time can be decreased by increasing the number of initial
state photons in the cavity. The increase in cavity decay probability with
increase in photon number is likely to reduce the $|W_{1}\rangle $ state
generation probability and must be carefully accounted for.

For the initial state preparation with $m+1=2$ and $n+1\geq 2$ the state of
the system at current time is%
\begin{align}
\Psi _{2,n+1}(t)& =\frac{e^{-i\omega _{1}t}}{(5n+3)}\left( (3n+3)\cos {%
\left( \sqrt{2(5n+3)}g\eta t\right) }+2n\right) |000\rangle \otimes
|2,n+1\rangle  \notag \\
& -ie^{-i\omega _{1}t}\sqrt{\frac{(3n+3)}{(5n+3)}}\sin {\left( \sqrt{2(5n+3)}%
g\eta t\right) }|W_{1}\rangle \otimes |1,n\rangle  \notag \\
& +\frac{e^{-i\omega _{1}t}\sqrt{6n(n+1)}}{(5n+3)}\left( \cos {\left( \sqrt{%
2(5n+3)}g\eta t\right) }-1\right) |W_{2}\rangle \otimes |0,n-1\rangle .
\label{19}
\end{align}%
The period for the system to return to initial separable state is $t_{p}=%
\frac{2\pi }{g\eta \sqrt{2(5n+3)}}$. At $t_{w}=\frac{\pi }{g\eta \sqrt{%
2(5n+3)}},$ we have ions in $W_{2}$-like state coupled to photon-phonon
system in state%
\begin{equation}
\Psi _{2,n+1}(t)=\frac{e^{-i\omega _{1}t}(n+3)}{(5n+3)}|000\rangle \otimes
|2,n+1\rangle +\frac{e^{-i\omega _{1}t}2\sqrt{6n(n+1)}}{(5n+3)}|W_{2}\rangle
\otimes |0,n-1\rangle .
\end{equation}%
We notice that the maximum probability of finding the state $|W_{2}\rangle $
increases with $n$, approaching $\cong \frac{24}{25}$ in the large $n$
limit. When initial state phonon and photon number is greater or equal to
three, all the four coupled basis vector states seen in eq. (\ref{13}) can
be reached.

\subsection{W- state Generation probabilities}

Three qubit W-states are extremely useful for implementing various
communication protocols, quantum state transport and quantum gates. These
states do not have genuine tripartite entanglement, but maximal pairwise
entanglement. For these states, global negativity and partial two way
negativity is non zero, while the partial three way negativity is zero \cite%
{shar082}, which is equivalent to the three tangle \cite{coff00} being zero.
The tripartite entanglement of the state is due bipartite correlations. {We
label the three ions as subsystems }${A}${, }${B}${, }${C}${, and consider
the phonons and photons to constitute a single quantum system }$D${.\ }The
reduced state operator for the ions 
\begin{equation*}
\rho _{m+1n+1}^{ABC}(t)=tr_{D}\left( \left\vert \Psi
_{m+1,n+1}(t)\right\rangle \left\langle \Psi _{m+1,n+1}(t)\right\vert
\right) 
\end{equation*}%
obtained by tracing over the vibrational and cavity state degrees of freedom
is a mixed state given by 
\begin{equation}
\rho _{m+1n+1}^{ABC}(t)=\left\vert a_{0}(t)\right\vert ^{2}|000\rangle
\langle 000|+\left\vert a_{1}(t)\right\vert ^{2}|W_{1}\rangle \langle
W_{1}|+\left\vert a_{2}(t)\right\vert ^{2}|W_{2}\rangle \langle
W_{2}|+\left\vert a_{3}(t)\right\vert ^{2}|111\rangle \langle 111|.
\label{redions}
\end{equation}%
The probability\text{ }$P_{1}(\tau )$ and $P_{2}(\tau )$ in figure (\ref%
{fig1}), are the probabilities of finding the three ions in the state $%
|W_{1}\rangle $ and $|W_{2}\rangle $ for the choice $m+1=n+1=3$. We notice
that for $t=\frac{3\pi }{8{g\eta }}$, $P_{0}(\tau )=P_{2}(\tau )=0$, $%
P_{1}(\tau )$ shows a peak and $P_{3}(\tau )$ is finite.

The choice $m=n=0$, in Eq. (\ref{13}) yields%
\begin{equation*}
\rho _{11}^{ABC}(t)=\cos ^{2}\left( \sqrt{3}{g\eta t}\right) |000\rangle
\langle 000|+\sin ^{2}\left( \sqrt{{3}}{g\eta t}\right) |W_{1}\rangle
\langle W_{1}|,
\end{equation*}%
with deterministic $|W_{1}\rangle $ state generation at ${t=}\frac{k\pi }{2{%
g\eta }\sqrt{{3}}}$, where $k$ is an odd integer.

For the choice $m=n=1$, in Eq. (\ref{13}) we get%
\begin{align}
\rho _{22}^{ABC}(t)& =\left( \frac{3}{4}\cos \left( {4g\eta t}\right) +\frac{%
1}{4}\right) ^{2}|000\rangle \langle 000|+\frac{3}{4}\sin ^{2}\left( {4g\eta
t}\right) |W_{1}\rangle \langle W_{1}|  \notag \\
& +\frac{3}{16}\left( \cos \left( {4g\eta t}\right) -1\right)
)^{2}|W_{2}\rangle \langle W_{2}|.  \label{20}
\end{align}

For an interaction time of $t=\frac{k\pi }{8{g\eta }},k=1,3,5,...,$ the
ionic state $|W_{1}\rangle $ is found with a probability of $75\%$ in the
state 
\begin{equation}
\rho _{22}^{ABC}\left( \frac{k\pi }{8{g\eta }}\right) =\frac{1}{16}%
|000\rangle \langle 000|+\frac{3}{4}|W_{1}\rangle \langle W_{1}|+\frac{3}{16}%
|W_{2}\rangle \langle W_{2}|,  \label{21}
\end{equation}%
whereas the probability of finding the three ions in state $|W_{2}\rangle $
is maximized for $t=\frac{k\pi }{4{g\eta }},$ with the reduced state
operator reading as 
\begin{equation}
\rho _{22}^{ABC}\left( \frac{k\pi }{4{g\eta }}\right) =\frac{1}{4}%
|000\rangle \langle 000|+\frac{3}{4}|W_{2}\rangle \langle W_{2}|.\quad
\label{22}
\end{equation}%
For $t=\frac{k\pi }{4{g\eta }}$ with $k=0,2,4,...$ the three ions are found
in their ground states. The three ion state is a mixed state, $\frac{1}{3}%
|W_{1}\rangle \langle W_{1}|+\frac{2}{3}|W_{2}\rangle \langle W_{2}|$, for ${%
g\eta t}$ values such that $\cos \left( {4g\eta t}\right) =-\frac{1}{3}.$

\begin{figure}[t]
\centering \includegraphics[width=3.75in,height=5.0in,angle=-90]{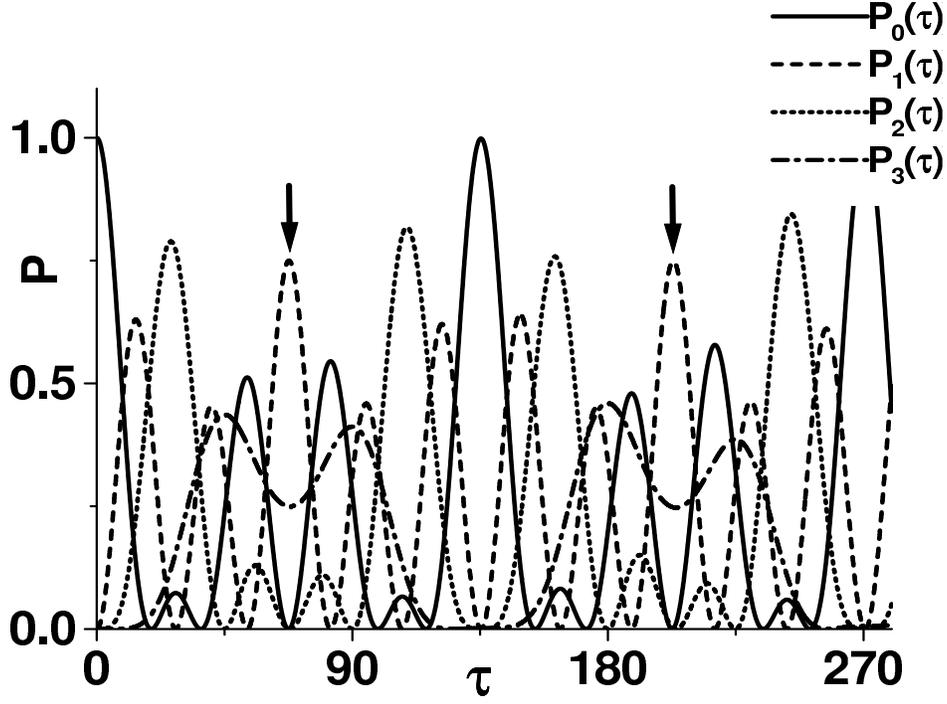}
\caption{The probabilities $P_{i}$, $i=0$ to $3$ versus $\protect\tau(=g 
\protect\eta t)$ for the initial state $|000,3,3 \rangle$.}
\label{fig1}
\end{figure}

\begin{figure}[t]
\centering \includegraphics[width=3.75in,height=5.0in,angle=-90]{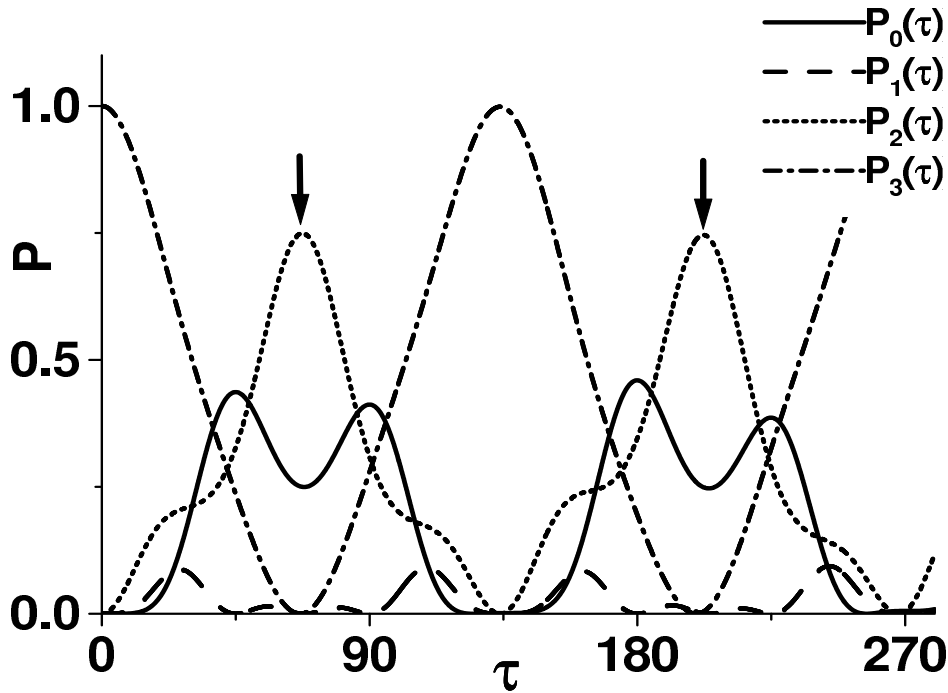}
\caption{The probabilities $P_{i}$, $i=0$ to $3$ versus $\protect\tau (=g%
\protect\eta t)$ for the initial state: $|111,0,0\rangle $.}
\label{fig2}
\end{figure}

The three ion state operator ($\rho _{m-2,n-2}^{ABC}(t)=tr_{D}\left(
\left\vert \Phi _{m-2,n-2}(t)\right\rangle \left\langle \Phi
_{m-2,n-2}(t)\right\vert \right) $), reads as%
\begin{equation}
\rho _{m-2,n-2}^{ABC}(t)=\left\vert a_{3}(t)\right\vert ^{2}|000\rangle
\langle 000|+\left\vert a_{2}(t)\right\vert ^{2}|W_{1}\rangle \langle
W_{1}|+\left\vert a_{1}(t)\right\vert ^{2}|W_{2}\rangle \langle
W_{2}|+\left\vert a_{0}(t)\right\vert ^{2}|111\rangle \langle 111|.
\end{equation}

For $m=n=2$, the $W_{2}$ state population probability at peak value is found
to be $75\%$ as seen in figure (2). The probability of populating $W_{1}$
state is, relatively, small.

\section{Entanglement of ionic qubits and Photons}

The entanglement of ionic memory qubits in W-like state to cavity photons is
an extremely interesting and useful aspect of the proposed scheme. Entangled
cavity photons can transport information to a remote cavity in a fast and
reliable way. {For analyzing the entanglement dynamics, we consider the }$%
(m+1-i)$ {phonons and (}$n+1-i)$ {photons to constitute a single quantum
system in the space spanned by vectors }$\left\vert {i}\right\rangle ${\ ($%
i=0$ to }${3)}$, where $\left\vert {i}\right\rangle $ represents the state{\ 
$|m+1-i,n+1-i\rangle $. Furthermore, the ground and excited state of an ion
represent logical bits }$\left\vert 0\right\rangle $ and $\left\vert
1\right\rangle ,$ respectively. The composite state of Eq. (\ref{13}), in
logical basis, reads as

\begin{eqnarray}
\Psi _{m+1,n+1}(t) &=&a_{0}(t)|0000\rangle +a_{1}(t)\left( \frac{%
|1001\rangle +|0101\rangle +|0011\rangle }{\sqrt{3}}\right)  \notag \\
&&+a_{2}(t)\left( \frac{|1102\rangle +|1012\rangle +|0112\rangle }{\sqrt{3}}%
\right) +a_{3}(t)|1113\rangle \text{.}  \label{25}
\end{eqnarray}%
Labelling the three ions as $A$, $B$, $C$ and cavity field plus phonon state
as subsystem $D$, the possible bipartite partitions of the system are $A-BCD$%
, $B-ACD$, $C-ABD$, $D-ABC$, $AB-CD$, $AC-BD$, and $AD-BC$. We notice that
the ionic state is symmetrical with respect to interchange of a pair of
qubits. As such distinct partitions are reduced to $A-BCD$, $D-ABC$, and $%
AB-CD$. The state is in bi-orthogonal Schmidt form for qubits $A$, $B$, $C$
and subsystem $D$.\ Quantifying or even detecting the entanglement of the
composite system with four subsystems is a fairly complex task. The four
subsystems may have four-partite GHZ state like correlations, or $4$-partite
entanglement resulting from bipartite entanglement between the subsystems.
On tracing over anyone of the subsystems, $4$-party GHZ like entanglement is
destroyed, but the three remaining subsystems may have tripartite GHZ like
entanglement or W-like entanglement. While the GHZ-like tripartite
entanglement is completely destroyed on tracing over one subsystem out of
the three, the reduced bipartite mixed state may still have bipartite
entanglement (provided that the tripartite system had W-like entanglement).
The entanglement distribution in the pure state of the composite system
determines the entanglement available simultaneously to all the four
parties, three selected parties or a pair of subsystems. In particular, the
nature, distribution and dynamics of entanglement involving the memory
qubits (trapped ions $ABC$) and photons is of special interest if the system
is part of a quantum network.

A. Peres \cite{pere96} was the first to observe that a partial transpose of
density matrix associated with a separable state of a bipartite system is
still a valid density matrix and thus positive (semi) definite. Horodecki et
al. \cite{horo96} proved that a positive partial transpose (PPT) was a
necessary and sufficient condition for separability of a state if the
dimension of the Hilbert space does not exceed six. A negative partial
transpose is a clear signature of entanglement. Negativity \cite%
{zycz98,vida00,eise01} based on Peres Horodecki PPT criterion has been shown
to be an entanglement monotone \cite{vida02}. In a recent article \cite%
{shar07}, we introduced $2-$way and $3-$way negativities to discuss the
entanglement of three qubit states. In refs. \cite{shar06b, shar08}, a
characterization of multipartite quantum states having $N$ subsystems, based
on partial $K-$way negativities has been proposed. The $K-$way partial
transpose with respect to a subsystem is constructed from the state operator
by imposing specific constraints on the matrix elements involving the states
of \ $K$ subsystems of multipartite composite system. It has been shown that
the partial transpose of density matrix of an N-partite system, with respect
to a given subsystem, can be written as a sum of $K-$way partial transposes
where $2\leq K\leq N$. Contribution to negativity due to a $K-$way partial
transpose is easily calculated. The $K-$way negativity ($2\leq K\leq N)$,
defined as the negativity of $K-$way partial transpose, quantifies the $K-$%
way coherences of the composite system. In contrast with the entropic
measures of entanglement, where a reduced state operator is used to obtain
information about the correlations between parts of a composite system, the\
partial $K-$way negativities are calculated from the state operator of the
composite system itself. The negativity of global partial transpose remains
invariant under local operations and classical communication (LOCC),
however, the partial $K-$way negativities may increase or decrease under
LOCC at the cost of each other. The partial $K-$way negativities show the
entanglement distribution in different parts of the system in a given state.
While the total entanglement in a composite system can not be increased by
local operations, the entanglement distribution can be changed by local
operations. In this section we investigate the entanglement dynamics of the
composite system state by using the global negativity to detect the
entanglement, and partial $K-$way negativities to determine the entanglement
distribution amongst different subsystems.

\subsection{Definition of Global and $K-$way negativities}

The global partial transpose $\widehat{\rho }_{G}^{T_{A}}$ of four-party
state operator $\widehat{\rho }^{ABCD}=$ $\left\vert \Psi \right\rangle
\left\langle \Psi \right\vert $ with respect to subsystem $A$ is constructed
by using the condition%
\begin{equation}
\left\langle i_{1}i_{2}i_{3}i_{4}\right\vert \widehat{\rho }%
_{G}^{T_{A}}\left\vert j_{1}j_{2}j_{3}j_{4}\right\rangle =\left\langle
j_{1}i_{2}i_{3}i_{4}\right\vert \widehat{\rho }^{ABCD}\left\vert
i_{1}j_{2}j_{3}j_{4}\right\rangle \text{.}  \label{GT}
\end{equation}%
The $K-$way partial transpose ($2\leq K$ $\leq 4$) of state operator $%
\widehat{\rho }^{ABCD}$ with respect to subsystem $A$ is defined as 
\begin{equation}
\left\langle i_{1}i_{2}i_{3}i_{4}\right\vert \widehat{\rho }%
_{K}^{T_{A}}\left\vert j_{1}j_{2}j_{3}j_{4}\right\rangle =\left\langle
j_{1}i_{2}i_{3}i_{4}\right\vert \widehat{\rho }^{ABCD}\left\vert
i_{1}j_{2}j_{3}j_{4}\right\rangle \quad \text{if}\quad
\sum\limits_{m=1}^{N}(1-\delta _{i_{m},j_{m}})=K,  \label{KT}
\end{equation}%
and 
\begin{equation*}
\left\langle i_{1}i_{2}i_{3}i_{4}\right\vert \widehat{\rho }%
_{K}^{T_{A}}\left\vert j_{1}j_{2}j_{3}j_{4}\right\rangle =\left\langle
i_{1}i_{2}i_{3}i_{4}\right\vert \widehat{\rho }^{ABCD}\left\vert
j_{1}j_{2}j_{3}j_{4}\right\rangle \quad \text{if}\quad
\sum\limits_{m=1}^{N}(1-\delta _{i_{m},j_{m}})\neq K,
\end{equation*}%
where 
\begin{equation}
\left[ 
\begin{array}{c}
\delta _{i_{m},j_{m}}=1\text{ for }i_{m}=j_{m} \\ 
\delta _{i_{m},j_{m}}=0\text{ for }i_{m}\neq j_{m}%
\end{array}%
\right. .
\end{equation}%
Similar constraints are applied to construct global and $K-$way partial
transpose with respect to subsystems $B$, $C$ or $D$. The negativity of $%
\rho ^{T_{p}}$ is related to the trace norm of $\rho ^{T_{p}}$ through%
\begin{equation}
N^{p}=\frac{1}{d_{p}-1}\left( \left\Vert \rho ^{T_{p}}\right\Vert
_{1}-1\right) .
\end{equation}

Negativity based on Peres Horodecki criterion is a natural entanglement
measure. We define the global negativity with respect to subsystem $p$ as 
\begin{equation}
N_{G}^{p}=\frac{1}{d_{p}-1}\left( \left\Vert \rho _{G}^{T_{p}}\right\Vert
_{1}-1\right) ,
\end{equation}%
and $K-$way negativity \cite{shar06b, shar08} as 
\begin{equation}
N_{K}^{p}=\frac{1}{d_{p}-1}\left( \left\Vert \rho _{K}^{T_{p}}\right\Vert
_{1}-1\right) ,
\end{equation}%
where $d_{p}$ is the dimension of the Hilbert space associated with
subsystem $p$.

Furthermore, we define $3-$way partial transpose $\widehat{\rho }%
_{3}^{T_{A-ABC}}$ involving the qubits $ABC$\ as%
\begin{eqnarray}
\left\langle i_{1}i_{2}i_{3}i_{4}\right\vert \widehat{\rho }%
_{3}^{T_{A-ABC}}\left\vert j_{1}j_{2}j_{3}j_{4}\right\rangle &=&\left\langle
j_{1}i_{2}i_{3}i_{4}\right\vert \widehat{\rho }^{ABCD}\left\vert
i_{1}j_{2}j_{3}j_{4}\right\rangle \quad  \notag \\
\text{if}\quad \sum\limits_{m=1}^{4}(1-\delta _{i_{m},j_{m}}) &=&3,\text{and 
}i_{1}\neq j_{1}, \\
\left\langle i_{1}i_{2}i_{3}i_{4}\right\vert \widehat{\rho }%
_{K}^{T_{A-ABC}}\left\vert j_{1}j_{2}j_{3}j_{4}\right\rangle &=&\left\langle
i_{1}i_{2}i_{3}i_{4}\right\vert \widehat{\rho }^{ABCD}\left\vert
j_{1}j_{2}j_{3}j_{4}\right\rangle \quad  \notag \\
\text{if}\quad \sum\limits_{m=1}^{4}(1-\delta _{i_{m},j_{m}}) &=&3,\text{and 
}i_{1}=j_{1},
\end{eqnarray}%
and 
\begin{equation}
\left\langle i_{1}i_{2}i_{3}i_{4}\right\vert \widehat{\rho }%
_{K}^{T_{A-ABC}}\left\vert j_{1}j_{2}j_{3}j_{4}\right\rangle =\left\langle
i_{1}i_{2}i_{3}i_{4}\right\vert \widehat{\rho }^{ABCD}\left\vert
j_{1}j_{2}j_{3}j_{4}\right\rangle \quad \text{if}\quad
\sum\limits_{m=1}^{4}(1-\delta _{i_{m},j_{m}})\neq 3,
\end{equation}%
with the corresponding negativity%
\begin{equation}
N_{3}^{A-ABC}=\frac{1}{d_{p}-1}\left( \left\Vert \widehat{\rho }%
_{3}^{T_{A-ABC}}\right\Vert _{1}-1\right) .
\end{equation}%
Analogous definitions hold for the $3-$way partial transposes $\widehat{\rho 
}_{3}^{T_{A-ABD}}$, $\widehat{\rho }_{3}^{T_{A-ACD}}$, and $\widehat{\rho }%
_{3}^{T_{D-BCD}}$.

\subsection{Global and partial $K-$way negativity}

Global negativity with respect to a subsystem can be written as a sum of
partial $K-$way negativities. The global transpose with respect to subsystem 
$p$, written in its eigen basis is given by 
\begin{equation}
\widehat{\rho }_{G}^{T_{p}}=\sum\limits_{i}\lambda _{i}^{G+}\left\vert \Psi
_{i}^{G+}\right\rangle \left\langle \Psi _{i}^{G+}\right\vert
+\sum\limits_{i}\lambda _{i}^{G-}\left\vert \Psi _{i}^{G-}\right\rangle
\left\langle \Psi _{i}^{G-}\right\vert ,  \label{1n}
\end{equation}%
where $\lambda _{i}^{G+}$and $\left\vert \Psi _{i}^{G+}\right\rangle $ ($%
\lambda _{i}^{G-}$and $\left\vert \Psi _{i}^{G-}\right\rangle $) are the
positive (negative) eigenvalues and eigenvectors of $\widehat{\rho }%
_{G}^{T_{p}}$, respectively. Using the definition of trace norm and $%
Tr\left( \widehat{\rho }_{G}^{T_{p}}\right) =1,$ the negativity of $\widehat{%
\rho }_{G}^{T_{p}}$ is given by 
\begin{equation}
{N}_{G}^{p}=-\frac{2}{d_{p}-1}\sum\limits_{i}\left\langle \Psi
_{i}^{G-}\right\vert \widehat{\rho }_{G}^{T_{p}}\left\vert \Psi
_{i}^{G-}\right\rangle =-\frac{2}{d_{p}-1}\sum\limits_{i}\lambda _{i}^{G-}%
\text{.}  \label{2n}
\end{equation}%
The global transpose with respect to subsystem $p$, may also be rewritten as 
\begin{equation}
\widehat{\rho }_{G}^{T_{p}}=\sum\limits_{K=2}^{N}\widehat{\rho }%
_{K}^{T_{p}}-(N-2)\widehat{\rho }.  \label{3n}
\end{equation}%
Substituting Eq. (\ref{3n}) in Eq. (\ref{2n}), we get%
\begin{equation}
-2\sum\limits_{i}\lambda
_{i}^{G-}=-2\sum\limits_{K=2}^{N}\sum\limits_{i}\left\langle \Psi
_{i}^{G-}\right\vert \widehat{\rho }_{K}^{T_{p}}\left\vert \Psi
_{i}^{G-}\right\rangle +2(N-2)\sum\limits_{i}\left\langle \Psi
_{i}^{G-}\right\vert \widehat{\rho }\left\vert \Psi _{i}^{G-}\right\rangle .
\end{equation}%
Defining partial $K-$way negativity $E_{K}^{p}$ ($K=2$ to $N$) as%
\begin{equation}
E_{K}^{p}=-\frac{2}{d_{p}-1}\sum\limits_{i}\left\langle \Psi
_{i}^{G-}\right\vert \widehat{\rho }_{K}^{T_{p}}\left\vert \Psi
_{i}^{G-}\right\rangle \text{,}
\end{equation}%
and%
\begin{equation}
E_{0}^{p}=-\frac{2(N-2)}{d_{p}-1}\sum\limits_{i}\left\langle \Psi
_{i}^{G-}\right\vert \widehat{\rho }\left\vert \Psi _{i}^{G-}\right\rangle
\end{equation}%
we may split the global negativity for qubit $p$ as%
\begin{equation}
N_{G}^{p}=\sum\limits_{K=2}^{N}E_{K}^{p}-E_{0}^{p}\text{.}
\end{equation}%
To obtain tripartite GHZ state like correlations between three subsystems,\
we calculate 
\begin{equation}
E_{3}^{A-ABC}=-\frac{2}{d_{p}-1}\sum\limits_{i}\left\langle \Psi
_{i}^{G-}\right\vert \widehat{\rho }_{3}^{T_{A-ABC}}\left\vert \Psi
_{i}^{G-}\right\rangle ,
\end{equation}%
\begin{equation}
E_{3}^{A-ABD}=-\frac{2}{d_{p}-1}\sum\limits_{i}\left\langle \Psi
_{i}^{G-}\right\vert \widehat{\rho }_{3}^{T_{A-ABD}}\left\vert \Psi
_{i}^{G-}\right\rangle ,
\end{equation}%
and%
\begin{equation}
E_{3}^{A-ACD}=-\frac{2}{d_{p}-1}\sum\limits_{i}\left\langle \Psi
_{i}^{G-}\right\vert \widehat{\rho }_{3}^{T_{A-ACD}}\left\vert \Psi
_{i}^{G-}\right\rangle .
\end{equation}%
It is easily verified that for the system at hand $E_{3}^{A-ABC}=0$, and $%
E_{3}^{A-ACD}=E_{3}^{A-ABD}=E_{3}^{A}/2.0$. As such the three qubits have no
genuine tripartite entanglement.

\begin{figure}[t]
\centering \includegraphics[width=3.75in,height=5.0in,angle=-90]{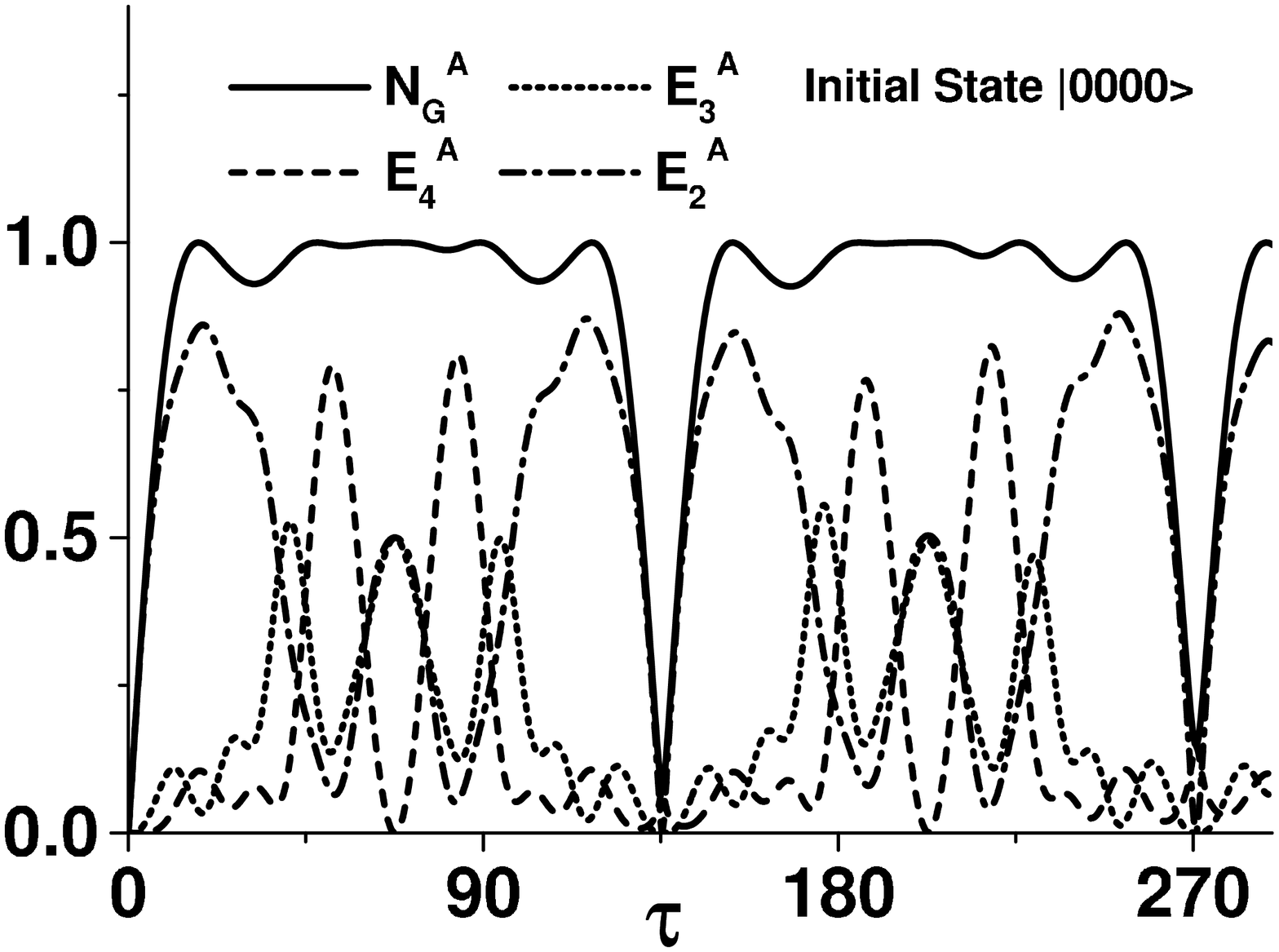}
\caption{The global negativity $N_{G}^{A}$, and entanglement measures $%
E_{K}^{A}$, for $K=2$ to $4$ versus $\protect\tau (=g\protect\eta t)$ for
the initial state $|0000\rangle $.}
\label{fig3}
\end{figure}

\begin{figure}[t]
\centering \includegraphics[width=3.75in,height=5.0in,angle=-90]{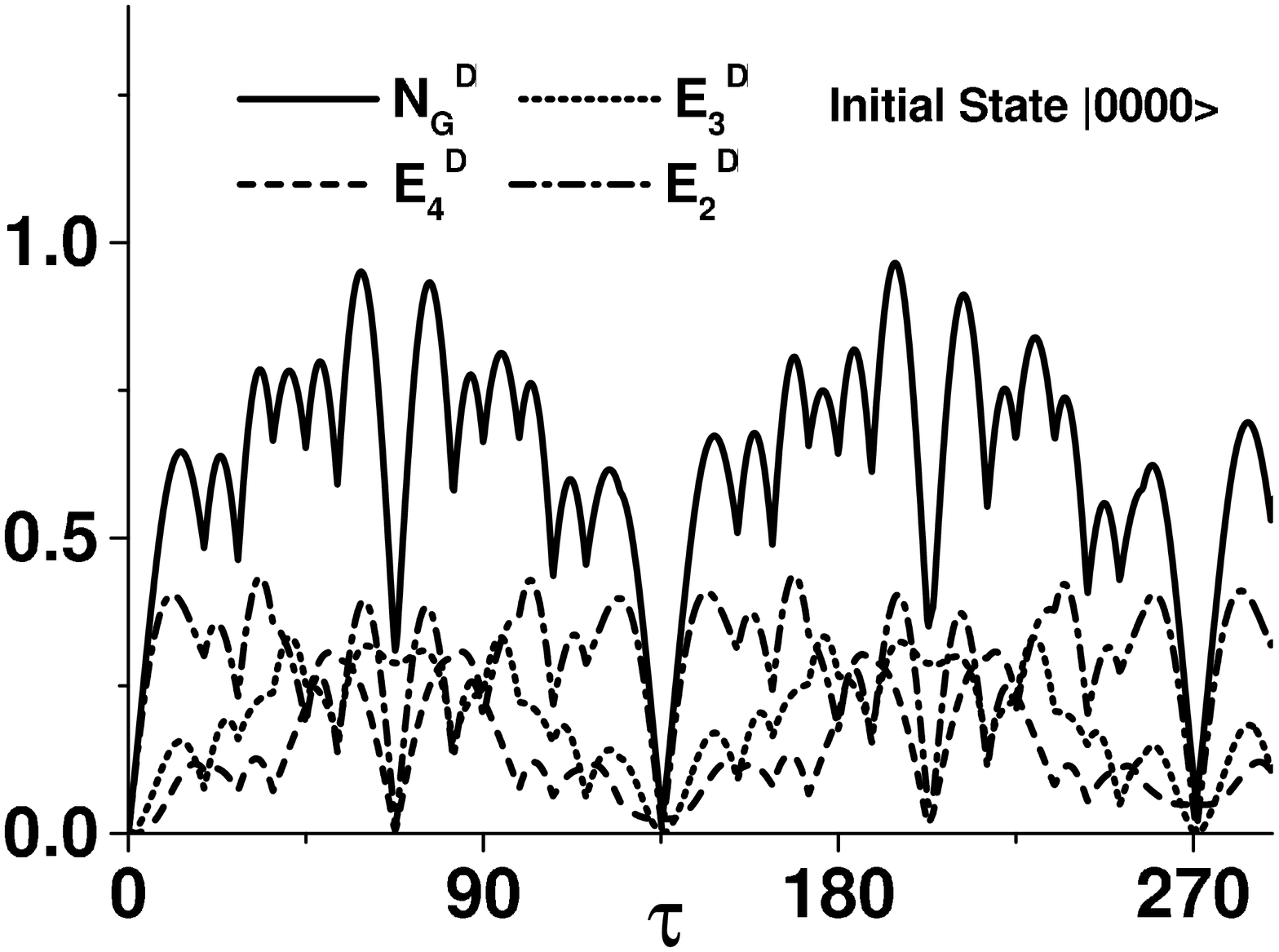}
\caption{The global negativity $N_{G}^{D}$, and entanglement measures $%
E_{K}^{D}$, for $K=2$ to $4$ versus $\protect\tau (=g\protect\eta t)$ for
the initial state $|0000\rangle $.}
\label{fig4}
\end{figure}

\begin{figure}[t]
\centering \includegraphics[width=3.75in,height=5.0in,angle=-90]{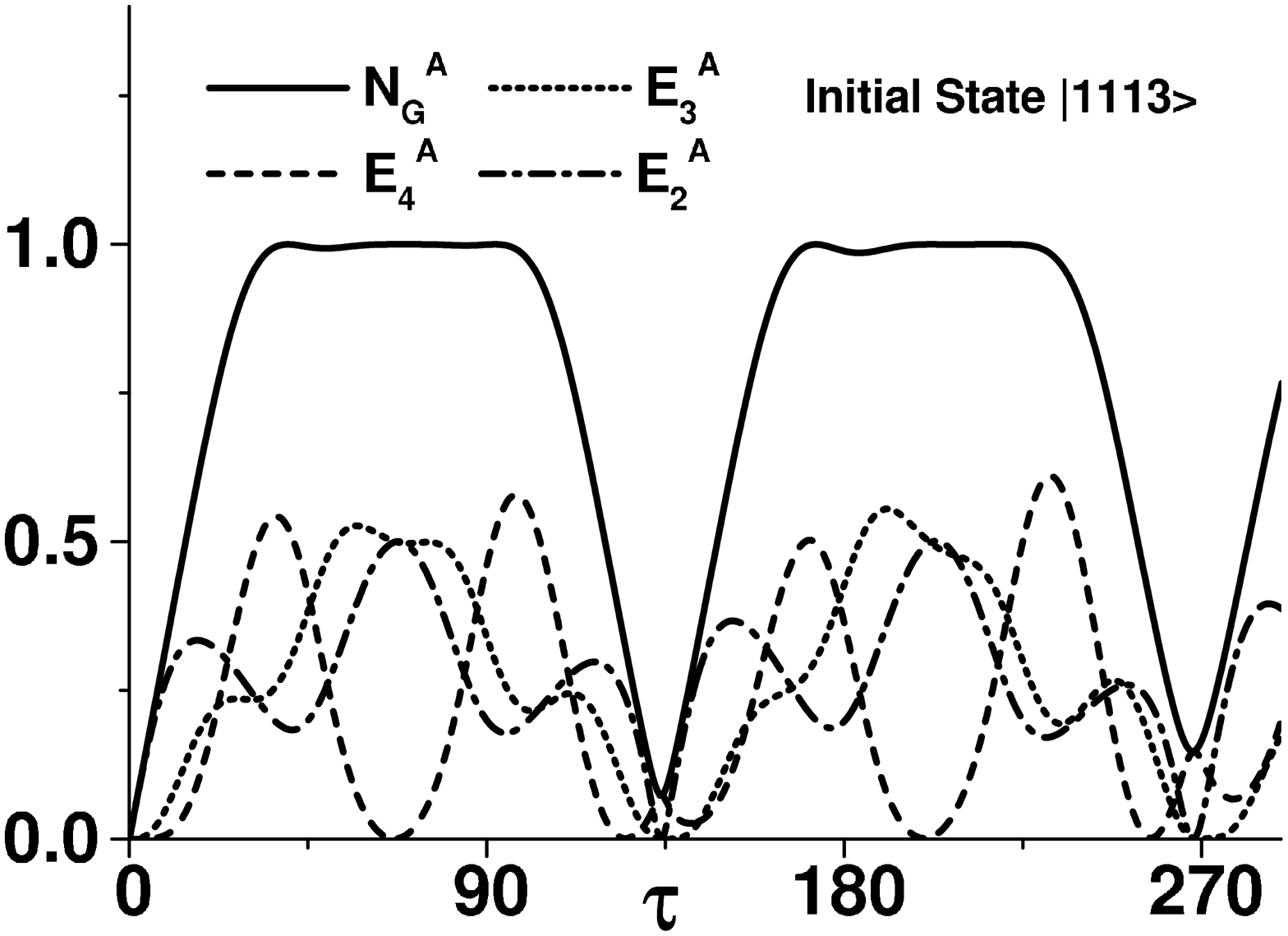}
\caption{The global negativity $N_{G}^{A}$, and entanglement measures $%
E_{K}^{A}$, for $K=2$ to $4$ versus $\protect\tau (=g\protect\eta t)$ for
the initial state $|1113\rangle $.}
\label{fig5}
\end{figure}

\begin{figure}[t]
\centering \includegraphics[width=3.75in,height=5.0in,angle=-90]{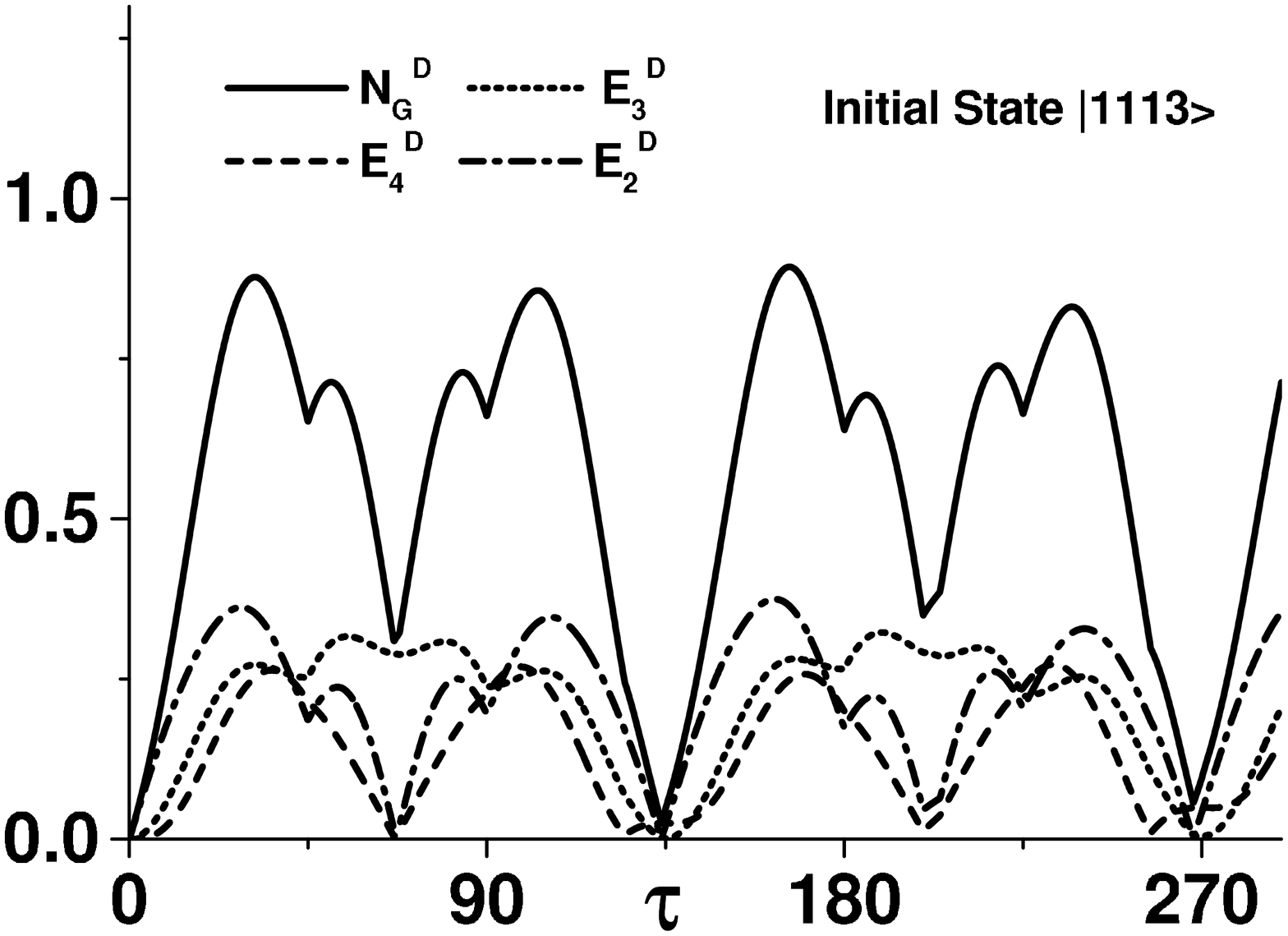}
\caption{The global negativity $N_{G}^{D}$, and entanglement measures $%
E_{K}^{D}$, for $K=2$ to $4$ versus $\protect\tau (=g\protect\eta t)$ for
the initial state $|1113\rangle $.}
\label{fig6}
\end{figure}

\begin{figure}[t]
\centering \includegraphics[width=3.75in,height=5.0in,angle=-90]{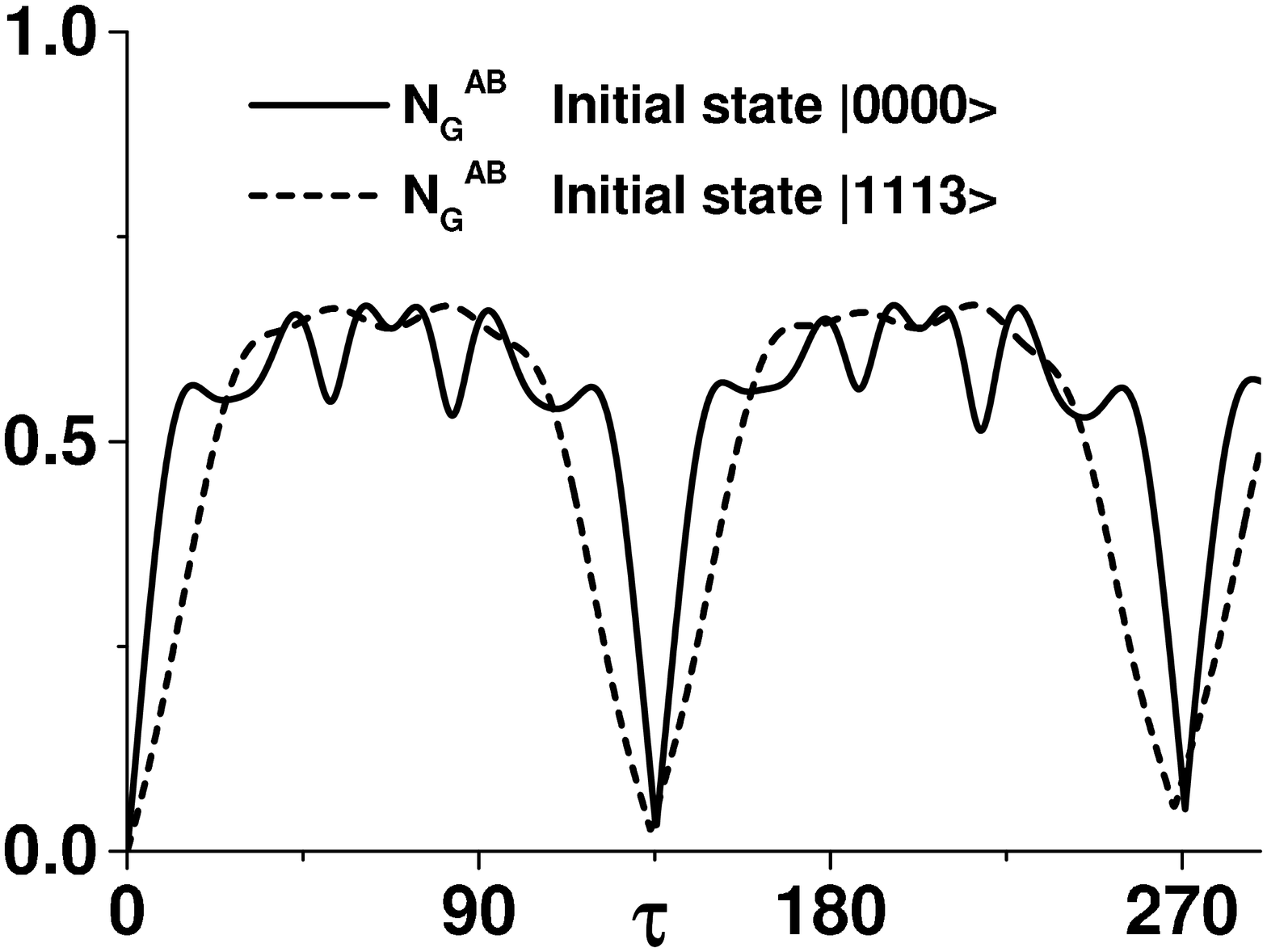}
\caption{The global negativity $N_{G}^{AB}$ versus $\protect\tau (=g\protect%
\eta t)$ for the initial states $|0000\rangle $ and $|1113\rangle $.}
\label{fig7}
\end{figure}

\subsection{Entanglement dynamics of the pure state $\Psi _{m+1,n+1}(t)$}

We use the global negativity to detect the entanglement of parts in
bipartite splits of the system. In case the negativities $N_{G}^{A},$ $%
N_{G}^{B},$ $N_{G}^{C},$ $N_{G}^{D}$ and $N_{G}^{AB}$ are non zero,\ the
system has 4-partite entanglement. The $4-$way, $3-$way and $2-$way partial
negativities identify and quantify the different types of entanglement
between the subsystems of the composite system in a given state. From the
state operator $\widehat{\rho }(t)=\left\vert \Psi
_{m+1,n+1}(t)\right\rangle \left\langle \Psi _{m+1,n+1}(t)\right\vert ,$ the
global and $K-$way partial transposes are constructed by following the
prescription given in section IV ( Eq. (\ref{GT}) and Eq. (\ref{KT})). The
state of Eq. (\ref{13}) is very special in that analytical results can be
obtained for partial $K-$way negativities characterizing the state. Using
negative eigen functions of $\rho _{G}^{T_{p}},$ the entanglement measures $%
E_{K}^{p}$ for $K=2,3$ and $4$ are easily obtained. The negativity of global
partial transpose with respect to qubit $A$ is found to be%
\begin{equation}
N_{G}^{A}=2\sqrt{\left( \left\vert a_{0}(t)\right\vert ^{2}+\frac{%
2\left\vert a_{1}(t)\right\vert ^{2}}{3}+\frac{\left\vert
a_{2}(t)\right\vert ^{2}}{3}\right) \left( \left\vert a_{3}(t)\right\vert
^{2}+\frac{\left\vert a_{1}(t)\right\vert ^{2}}{3}+\frac{2\left\vert
a_{2}(t)\right\vert ^{2}}{3}\right) }\text{.}  \label{27}
\end{equation}%
Using the eigenvector corresponding to the negative eigenvalue of $\rho
_{G}^{T_{A}}$, we get the partial negativities 
\begin{equation}
E_{4}^{A}=\frac{4}{N_{G}}\left( \left\vert a_{0}(t)\right\vert
^{2}\left\vert a_{3}(t)\right\vert ^{2}+\frac{\left\vert a_{1}(t)\right\vert
^{2}\left\vert a_{2}(t)\right\vert ^{2}}{3}\right) ,  \label{28}
\end{equation}%
\begin{equation}
E_{3}^{A}=\frac{4}{N_{G}}\left( \frac{2\left\vert a_{0}(t)\right\vert
^{2}\left\vert a_{2}(t)\right\vert ^{2}}{3}+\frac{2\left\vert
a_{1}(t)\right\vert ^{2}\left\vert a_{3}(t)\right\vert ^{2}}{3}\right) ,
\label{29}
\end{equation}%
and%
\begin{equation}
E_{2}^{A}=\frac{4}{N_{G}}\left[ \frac{\left\vert a_{0}(t)\right\vert
^{2}\left\vert a_{1}(t)\right\vert ^{2}}{3}+\frac{\left\vert
a_{2}(t)\right\vert ^{2}\left\vert a_{3}(t)\right\vert ^{2}}{3}+2\left( 
\frac{\left\vert a_{1}(t)\right\vert ^{2}}{3}+\frac{\left\vert
a_{2}(t)\right\vert ^{2}}{3}\right) ^{2}\right] .  \label{30}
\end{equation}%
It is easily seen that $N_{G}^{A}=N_{G}^{B}=N_{G}^{C}$. Next we construct
the transposes $\rho _{G}^{T_{D}}$, $\rho _{4}^{T_{D}}$, $\rho _{3}^{T_{D}}$%
, and $\rho _{2}^{T_{D}}$, for the decomposition $ABC-D$ of the state $\Psi
_{m+1,n+1}(t)$ and obtain 
\begin{eqnarray}
N_{G}^{D} &=&\frac{2}{3}\left\vert a_{0}(t)\right\vert \left( \left\vert
a_{1}(t)\right\vert +\left\vert a_{2}(t)\right\vert +\left\vert
a_{3}(t)\right\vert \right)  \notag \\
&&+\frac{2}{3}\left\vert a_{1}(t)\right\vert \left( \left\vert
a_{2}(t)\right\vert +\left\vert a_{3}(t)\right\vert \right) +2\left\vert
a_{2}(t)\right\vert \left\vert a_{3}(t)\right\vert \text{,}  \label{31}
\end{eqnarray}

\begin{equation}
\quad E_{4}^{D}=\frac{2}{3}\left\vert a_{0}(t)\right\vert \left\vert
a_{3}(t)\right\vert +\frac{2}{9}\left\vert a_{1}(t)\right\vert \left\vert
a_{2}(t)\right\vert ,
\end{equation}%
\begin{equation}
E_{3}^{D}=\frac{2}{3}\left\vert a_{0}(t)\right\vert \left\vert
a_{2}(t)\right\vert +\frac{2}{3}\left\vert a_{1}(t)\right\vert \left\vert
a_{3}(t)\right\vert ,
\end{equation}%
and%
\begin{equation}
E_{2}^{D}=\frac{2}{3}\left\vert a_{0}(t)\right\vert \left\vert
a_{1}(t)\right\vert +\frac{2}{3}\left\vert a_{2}(t)\right\vert \left\vert
a_{3}(t)\right\vert +\frac{4}{9}\left\vert a_{1}(t)\right\vert \left\vert
a_{2}(t)\right\vert .
\end{equation}%
Since the three qubits have no genuine tripartite entanglement, the partial
negativity $E_{3}^{D}$ represents the entanglement of quantum system $D$
with the $W$ like states of three qubits.

Treating $AB$ as a single system in Hilbert space of dimension four, the
global negativity is found to be%
\begin{equation}
N_{G}^{AB}=\frac{2}{3}\left( \mu _{0}\mu _{1}+\mu _{0}\mu _{2}+\mu _{1}\mu
_{2}\right) ,
\end{equation}%
where%
\begin{eqnarray}
\mu _{0} &=&\sqrt{\left\vert a_{0}(t)\right\vert ^{2}+\frac{\left\vert
a_{1}(t)\right\vert ^{2}}{3}},\quad \mu _{1}=\sqrt{\frac{2\left\vert
a_{1}(t)\right\vert ^{2}}{3}+\frac{2\left\vert a_{2}(t)\right\vert ^{2}}{3}},
\notag \\
\mu _{2} &=&\sqrt{\left\vert a_{3}(t)\right\vert ^{2}+\frac{\left\vert
a_{2}(t)\right\vert ^{2}}{3}}.
\end{eqnarray}

Figures (3) and (4) display the global negativity and $K-$way entanglement
of qubits $A$ and $D$ in the state $\Psi _{33}(t),$ as a function of
interaction parameter $\tau =g\eta t$. At $\tau \approx \frac{3\pi }{8}$, we
have $E_{3}^{A}=E_{2}^{A}=0.5$, indicating that the qubit $A$ (or $B$ or $C)$%
, has equally strong bipartite and tripartite correlations. Recalling that
no genuine tripartite entanglement exists between the three qubits as
evidenced by $E_{3}^{A-ABC}=0$, we have here the three qubits in $W_{1}$%
-like state entangled to the subsystem $D$. In other words, the cavity field
is entangled to three ions in $W_{1}$-like state and can be used to transfer the
entanglement of the composite system to a remote quantum system. The
probability plot of figure (\ref{fig1}) confirms that for interaction time $%
t=\frac{3\pi }{8{g\eta }}$, $P_{0}(\tau )=P_{2}(\tau )=0$, $P_{1}(\tau )$
shows a peak and $P_{3}(\tau )$ is finite. A measurement that finds the
cavity in two photon state collapses the composite system state to three
ions in $W_{1}$-state with center of mass in two phonon state.

\subsection{Entanglement dynamics of the pure state $\Phi _{m-2,n-2}(t)$}

Analytical expressions for negativities and partial $K-$way negativities for
the state $\Phi _{m-2,n-2}(t)$ (Eq. (\ref{23}))$,$ written in computational
basis as 
\begin{eqnarray}
\Phi _{m-2,n-2}(t) &=&a_{3}(t)|0000\rangle +a_{2}(t)\left( \frac{%
|1001\rangle +|0101\rangle +|0011\rangle }{\sqrt{3}}\right)  \notag \\
&&+a_{1}(t)\left( \frac{|1102\rangle +|1012\rangle +|0112\rangle }{\sqrt{3}}%
\right) +a_{0}(t)|1113\rangle \text{.}
\end{eqnarray}%
are analogous to those for the state $\Psi _{m+1,n+1}(t)$. For the special
case of state $\Phi _{00}(t),$ the global and partial $K-$way negativities
are displayed, in Figs. (5) and (6) for qubits $A$ and $D$, respectively.
The negativities $N_{G}^{AB}$ for the states $\Psi _{33}(t)$ and $\Phi
_{00}(t)$ are shown in Fig. (7). At $\tau =\frac{3\pi }{8}$, the coefficient 
$a_{2}(t)=a_{0}(t)=0$ , and $E_{3}^{A}=E_{2}^{A}=0.5$ giving (using Eq. \ref%
{psiw2}) the state%
\begin{equation}
\Phi _{00}(\tau =\frac{3\pi }{8})=a_{3}(t)|000,m+1,n+1\rangle
+a_{1}(t)|W_{2},m-1,n-1\rangle ,  \notag
\end{equation}%
In this case we have three ions in $W_{2}$-like state entangled to
photon-phonon system. As such detecting the cavity in single photon state
ensures that the three ions are in $W_{2}$ state. The entanglement of the
cavity field with three ions in $W_{2}$-like state may, on the other hand,
be used to communicate with a remote quantum system. For an interaction time 
$t=\frac{3\pi }{8{g\eta }}$, $P_{2}(\tau )$ shows a peak and $P_{0}(\tau )$
is finite in the the probability plot of figure (\ref{fig2}).

The number of initial state vibrational quanta also controls the nature of
entanglement in the composite system states. We notice that for a single
phonon initial state the general form of the state (Eq. (\ref{18})) 
\begin{equation}
\Psi _{1,n+1}(t)=a_{0}(t)|0000\rangle +a_{1}(t)\left( \frac{|1001\rangle
+|0101\rangle +|0011\rangle }{\sqrt{3}}\right) ,
\end{equation}%
allows only bipartite entanglement. For m $\geq $ $2$, the composite system
may have genuine 4-partite, tripartite as well as bipartite entanglement.
This is an interesting aspect unique to systems where vibrational motion of
ions is coupled to cavity field.

\section{What do the $K-$way negativities add to our knowledge of the system?%
}

Figures (3-7) show the entanglement distribution between possible
entanglement modes available to the system in states $\Psi _{33}(t)$\ and $%
\Phi _{00}(t)$. Since analytical expressions for partial $K-$way
negativities are available for specific initial state preparations, the rate
of change of partial $K-$way negativity for a particular mode can be easily
obtained, if needed. The two way partial negativity is seen to grow at the
fastest rate being the first to reach it's peak value, followed by two peaks
showing maxima of partial $3-$way and $4-$way negativities, respectively.
Besides that a reversible entanglement exchange between different
entanglement modes is observed. In Fig. (3) at the maxima of partial $4-$way
negativity, partial $2-$way and $3-$way negativities are rather small,
whereas the minima of $E_{4}^{A}$ correspond to a large contribution to
total entanglement from bipartite and genuine tripartite entanglement.
Similar trend is seen in Figures (4-6). Figure (7) complements the
information about entanglement distribution obtained from Figures (1-6).
Once the interaction is switched on, a typical qubit pair is found to be in
an entangled state until the composite becomes separable again. As seen from
the global negativity plots, the period after which the composite system
becomes separable is given by $T=\frac{3\pi }{4{g\eta }}.$

In particular, we look at the entanglement of states $\Psi
_{m+1,n+1}^{W_{1}} $ (Eq. (\ref{psiw1})) and $\Phi _{m-2,n-2}^{W_{2}}$ (Eq. (%
\ref{psiw2})). For both types of states $N_{G}^{D}=E_{3}^{D}$ , that is
there no genuine four partite entanglement amongst the subsystems $A$, $B$, $%
C$ and $D$. The global negativities are however finite for $A$, $B$, $C$, $D$
as well as $AB, $ pointing to four partite entanglement. Although for qubits 
$A$, $B$, and $C $ we get finite partial $2-$way negativities, four partite
entanglement cannot be due to two-way correlations because $E_{2}^{D}=0.$
The $2-$way negativity of the partial transpose of $W_{1}$ state or $W_{2}$
state is $0.94$, where as the three way negativity is zero. As such a $W$
state is a state with maximal tripartite entanglement generated by $2-$way
correlations. In states $\Psi _{m+1,n+1}^{W_{1}}$ and $\Phi
_{m-2,n-2}^{W_{2}}$ ions in a $W$ like state are entangled to subsystem $D$.
We conclude that the four-partite entanglement of three ions with
photon-phonon system is generated by $2-$way and $3-$way correlations.

\section{ Conclusions}

We have studied the entanglement dynamics of spatially separated three
two-level cold trapped ions in a high finesse cavity with the cavity tuned
to the red sideband of ionic vibrational motion. Analytical expressions for
the state of composite system as a function of interaction time are obtained
for interaction Hamiltonian of Eq. \ref{3} with the cavity prepared
initially in different photon number states. With three ions initially in
their ground state, the state $|000,m+1,n+1\rangle $ evolves (i) for $m=0$
and $n\geq 0,$ in a bi-dimensional subspace , (ii) for $m=1$ and $n\geq 1$
in a tridimensional subspace, and (iii) for $m\geq 2$ and $n\geq 2$ in a
four-dimensional subspace of the coupled basis ionic states. The initial
state $|111,m+1,n+1\rangle $ always evolves in a four-dimensional subspace
independent of the initial phonon and photon \ number. The number of initial
state vibrational quanta offers a control mechanism for manipulation of
composite system states, when ions are in their ground state initially. This
is an interesting aspect unique to systems where vibrational motion of ions
is coupled to cavity field in contrast to the ions coupled only to quantized
cavity field \cite{fuji04}, \cite{cai05} or only with the vibrational modes 
\cite{retz06}. Useful practical applications to implement information
processing and communication related tasks, can benefit from this special
feature. The reduced three ion state operator is obtained by tracing out the
phonon and photon degrees of freedom. {For cavity ion coupling strength }$%
g=8.95$ MHz, Lamb Dicke parameter value $\eta =0.01$ and {cavity prepared in
single photon state at }$t=0$, the minimum interaction time needed generate
a three ion W-state is found to be $\sim 10.133\mu $ sec. For the initial
state in which three ions are in their ground state and center of mass in
two phonon state, the $W_{2}$ state generation probability increases while
the minimum interaction time to get the probability peak decreases, with
increase in the number of photons present in the cavity at $t=0$.

The ionic qubits in W state are found to be entangled to cavity photons,
that may be used to transport information to a remote cavity in a fast and
reliable way. Multipartite entanglement dynamics of the composite system is
examined using global, four-way, three-way and two-way negativities.
Analytical expressions for partial $K-$way negativities\ ($K=2$ to $4$) are
obtained. For the three ions prepared initially in their ground state or in
their excited state, the partial $K-$way negativities are calculated
numerically and plotted as a function of interaction parameter. These plots
show the entanglement distribution as well as the rate of change with time
of entanglement between possible entanglement modes available to the system.
Besides that reversible entanglement exchange between different entanglement
modes is observed. For specific values of interaction parameter, the three
ions and photon-phonon system are found to have four partite entanglement,
generated by $2-$way and $3-$way correlations. Three ions in W state are
found to be entangled to\ photons. We expect this analysis to add to the
understanding of multipartite entanglement in the context of trapped ions
interacting with photons in optical cavities.

{\LARGE Acknowledgments}

S. S. Sharma acknowledges financial support from FAEP/UEL, and Funda\c{c}%
\~{a}o Araucaria PR, Brazil. E. de Almeida thanks Capes, Brazil for
financial support.

\end{document}